# The *Deviants' Dilemma*: A Simple Game in the Run-Up to Seniority

AKM Azhar [*,a], WAT Wan Abdullah, [**]  ZA Kidam [*], A. Salleh [*]

[*]Graduate School of Management
Universiti Putra Malaysia
43400 Serdang, Selangor
Malaysia

[**] Department of Physics
University of Malaya
50603 Kuala Lumpur
Malaysia

**Email:** [a] akmazhar@putra.upm.edu.my

**Abstract**

Seniority runs deep. As the saying goes for some conventions or institutions in some societies, seniority exhibits and engulfs into almost all nooks and corners of their organisations. This paper presents an analysis of the conflict between two groups in the run-up to seniority in these institutions. We overview the inspirations and practice of each group, and their conflict in the run-up to organisational seniority, and analyse the conflict using a simple model of a "Deviants' Dilemma" game played in a single population. Within the game's framework, we analyse the conditions for the evolutionary stable state and the stability of the proportions of each group, given a strategy resulting from learning based on rewards. We also provide a short discussion on the role of noise in the system and we conclude with some discussions on the generality and possible applications of the model.

---

# 1. INTRODUCTION

Quite sometimes in life, things do not work nicely in the favour of certain groups. Also sometimes if you're the loser, in spite of your patience, persistence and perseverance, you're still stuck with the outcome. Consider the cases of appointments to high positions or promotion exercises in some societies. There are things that first require cooperation from both parties, the promoting and the promoted. Let us consider an academic set-up, and continue to do so for the purposes of this paper. First on the eventual winner, he begins by projecting himself as the leader in his academic field in his institution so he stands a very good chance of being selected to head, say, a high profile research committee managing substantial amounts of funds. Or suppose that you have long aspired to be appointed as a professor in your university. You follow standard procedures, fill in all required forms, present your *curriculum vitae* highlighting the numbers, and presumably mask any academic benchmarks higher than the one that you use, and then enter yourself for promotion, which is to be decided by a panel committee whose members do not know you nor understand your field of specialization that well. Once your employer makes such a commitment and you get appointed, you start playing the *pseudo*'s game after enjoying the perks and privileges of the job. The university, the loser, is stuck – with a large salary obligation and you *merci beaucoup*, will be laughing all the way to the bank! These are just two of the many prevalent examples.

The university's headache may essentially be in the form of a one-shot game. Would the university's panel appoint you to the post unless they believe and understand that your professional experience and qualifications are commensurate with the appointment? This is one of the so-called "commitment dilemmas" of the university[1]. Once you have played the game nicely and convinced the interview panel (and the university top management is convinced) that you truly deserved to be appointed to the professorial post, this is where the fun steps in (for you of course!); you will then have many opportunities when it will be to your advantage to (academically) "cheat" on the trusted appointment.

The remainder of this paper proceeds as follows: In the next section, we provide an overview of the aspirations and practice of *deviants* and *pseudos*, two groups in a single population who upon instruction, encounter into a conflict. In Section 3, we formalize the encounter in a

---
[1] This so called "commitment dilemmas" was described beautifully by Robert Frank of Cornell University in [Frank, 1989].

**2**

simple evolutionary game model - we present the Deviants' Dilemma (DD) game. Within the framework of the assumed payoffs, we analyse the conditions for the evolutionary stable state and the stability of the proportions of each group. Section 4 extends the analysis briefly by considering the role of noise. In section 5, we conclude with some discussions on the possible applications of the model.

## 2. THE PLAYERS

Suppose that there may have been attempts in implementing some prerequisites in the quest for excellence in academic institutions. These "overnight" efforts may be sincere in the context of intentions and objectives. Academic meritocracy finally may soon be stamping its mark in campus promotions. But the problem is establishments do not come and go overnight and actual practices may not turn in favour towards those who persevere to meet international standards[2].

Consider the first group, *pseudos*, who strive to be deemed individuals of great merit, but are hardly arguably so. Once they are in the driver's seat, for any act of them continuing to play the pseudo's game, it is the system of the day which dictates as to whether they are safe or not. Most or all of the time pseudos will never be penalized for definitions of quality can be somewhat tuned to be relative[3].

Suppose then there is another group in the institution, a so-called *deviant* group who willingly impose the international benchmark upon them. Deviants claim there are lot of pseudos who "cheat" and prosper, rising their way up the academic ladder. They argue that it's common sight in campuses to have seniors flagging their seniority and asking juniors to *get on* academically but what juniors receive from seniors in return for all the high regards and respect instead are repeated displays and signals for the former to emulate the latter's trade. The pseudos' sermon is "one has to play by the *you can't change the system* rules, project yourself and work your way to be appointed in at least some administrative posts these are the so called *power speaks* stuff, look smart, look busy, commit yourself to attend lots and lots of meetings, and publish lots and lots of papers as opposed to a few deep ones"[4].

---

[2] Deviants claim there may have been many allegations of heterogeneous criteria being applied to different candidates with respect to publications in promotion exercises. This may in some probability be due to existing non-objective benchmarks in the promotion system.

[3] An example of the deviants argument on the relativity of quality: Suppose all academic staff members in *Bagus University* applying for PhD research study leave in the UK are required to gain admission into "5" ranked tertiary institutions as one of the conditions for the eligibility of their scholarship award. Following which, by consistency rules, *Bagus University* will also have to adapt the objective criteria utilised by the UK's Research Assessment Exercise (RAE) panel to arrive at their respective rankings of each institution into their merit system. A simple example of which is the list and the ranking of journals publications used by the RAE panel.

Deviants insist on the transparency and fairness of the promotion system and productivity and quality to be defined according to an objective benchmark[5]. The Deviants' conjecture is that cultural and traditional norms are still encapsulating their academic environment hence research and papers emancipating from research are all regarded as of the same quality - numbers, and pages will impress and will count heavily. Deviants say pseudos' confuse assessors with both rewarded and un-rewarded activities, and dummy their way through promotion and never *seem* to be bothered less caught. They claim there are many pseudo academicians who are quite expert at the above game, and their impression of pseudos' summarised by Gerard Silverberg's *"constantly hashing and rehashing the same fundamental material in lieu of genuinely new results"*.

Next, the story continues with sudden changes in the university's top management personnel. This new management started implementing new standards with regards to promotion. Following exposure and studies on "actual" international standards, supposedly, the institution's top management begin to recognise international benchmarks and start amending promotion regulations. Consider a new institutional ruling passed such that academicians do have to work in pairs, and output of this supposedly cooperation will be assessed and submitted to institutional authorities for annual evaluation and promotion considerations. Intra institution wise, academicians are required to pair up, cooperate in research, and apply for annual promotion to be decided by the institution's panel.

We now bring the above encounter and conflict of pseudos and deviants into analysis. Assume that neither deviant nor pseudo can tell whether a chosen working partner belongs to which group. What will then be the proportion of future academia? Furthermore, if the pseudo's practice is such an attractive strategy, why don't all deviants dump the international benchmark and conform to the pseudo game? This is the evolutionary problem for deviants. Once a deviant chooses to cooperate, and mistakenly ends up with a pseudo partner, in the reality of his operating environment, he'll be stuck with the outcome, too bad for him then. Will pseudo*s* be soon taking over these institutions? Will future academia be full of pseudos? Will pseudomania be running rampage in campus? We proceed to describe a simple model to answer some of these questions in the next section.

---

[4] What's the issue here? It's some of the Deviants Inspirations: Hal Varian of UC Berkeley (in *The AEA's Electronic Publishing Plans: A Progress Report*) suggested "universities adopt a policy of putting a limit on the number of papers they will accept for purposes of tenure review" (i.e. in the promotion exercise of an assistant professor to that of an associate). He stressed that "authors can focus themselves on doing a few serious pieces rather lots of shallow ones", and quoted Herb Simon's *"wealth of information creates a poverty of attention"*. What deviants say in context of Hal Varian's suggestion to their experience is that "we have truckloads of cases of wealth of information creating wealth of admiration!"

[5] The conflict arising between deviant and pseudo is because the former plays to the tune of the international benchmark, while the latter endeavours to uphold existing academic norms and standards – *repertoire* of traditional institutional practices.

## 3. THE MODEL

Consider the choice of deviant and pseudo being represented by a "Deviants' Dilemma" (DD) game, assuming two-person interactions as the basis for academic cooperation. Say a deviant gets payoff of $c$ when paired with a deviant, but 0 when stuck with a pseudo (deviants just cannot work with pseudos — they just turn disequilibrated and get nothing!). However, pseudos just love to pair up with deviants and gets $b$ ($b>c$) when he goes with a deviant. Of course pseudomania is unoriginal, they just would not want to work hard enough to generate new material, and hence they try to avoid one another, but if stuck each will get only $a$ ($a<c$). This is represented by the following symmetric 'fixed zero baseline' 2×2 payoff matrix:

| A \ B | Pseudo | Deviant |
|---|---|---|
| Pseudo | $a$ / $a$ | $b$ / 0 |
| Deviant | 0 / $b$ | $c$ / $c$ |

with the requirement that $0 < a < c < b$. Discounting a scaling factor, there are actually only two free parameters which define the system, which we represent as $\xi$ and $\eta$, where $\xi = b/a$ and $\eta = c/a$. Thus the requirement that $b > c$ translates as $\xi > \eta$.

Due to the choice of this payoff set-up, if there is no way that deviants can recognise pseudos, then the latter are going to predominate the academic population big time. Pseudos will then always do better, regardless of their proportion in academia. This is illustrated by the following one-shot analysis. Let's say the proportion of deviants in the university population is *p*, so that that of pseudos would be $(1-p) = \bar{p}$. Then assuming that the players exercise rational choice based on the allocated payoffs, the expected number of pseudos would always be greater than deviant for any value of *p*, since the average payoff for deviants is

$$pc + (1-p)0 = pc$$

while that for pseudos is

$$bp + (1-p)a.$$

One finds that payoffs for pseudos are always higher,

$$bp + (1-p)a > pc.$$

This is because $\xi > \eta$ and $\frac{(1-p)}{p} > 0$, yielding

$$\xi + \frac{(1-p)}{p} > \eta \quad \text{or} \quad \xi p + (1-p) > \eta p,$$

which leads to the inequality. Therefore in general, without the availability of information on potential partners, if we assume that proportion of the university's next set of incoming academician conforming themselves to either becoming a pseudo or deviant is based on the average score, pseudos tend to dominate for all values of *p* in this DD game.

Consider the case where information is involved in the system. Suppose a deviant can only identify a partner is a deviant or a pseudo by looking into the person's *curriculum vitae*.[6] Assume there's some cost to this effort. Let's say it costs them *r* or *a*ρ on average. Then a deviant can guarantee oneself a payoff of *c-r* points by ensuring he pairs up with another deviant. To be exact, the probability of finding a deviant after finding (*n*-1) pseudos is given by,

$$\Pr(n) = (1-p)^{n-1} p = \bar{p}^{n-1}(1-\bar{p})$$

giving the average number of verifications required before a deviant is found, to be

$$\bar{n} = \sum_{n=1}^{k} n \Pr(n) = (1-\bar{p}) \sum_{n=1}^{k} n \bar{p}^{n-1}.$$

where *k* is the number of available partnerships. However for simplicity we assume 1 is exactly the total cost; besides, for small values of $\bar{p}$ (i.e. large proportions of deviants), $\bar{n} = 1$ is a good approximation. If every deviant conducts this check before starting the cooperation, then pseudos would only have themselves to work with and get *a*, if so then deviant would soon take over the institutions and soon all universities will be full of deviants! The implicit point here is the finiteness of the system, where pseudos get *a* through exclusion, where otherwise they would get the highest payoff of *b* by also checking on their partners.

---

[6] More of deviants inspirations: They claim it may be common practice to see the majority of tertiary institutions both public and private highlighting and marketing their academic programmes but what is glaringly missing is for the public to view the academicians *curriculum vitae* (It seems that most academicians keep their objective *curriculum vitae* confidential except to the promotions panel for reasons they alone know best).

So what is the proportion *q* of deviants who will be paying the cost of checking on average? We have to find that value *q* that gives pseudos the same payoff on average as deviants. For the iterated game, assuming some kind of learning, the evolutionary stable state (ESS) equilibrium requirement is that the average payoff for deviant is the same, whether they check the identity of their working partner and pay the verification cost, or simply just take their chances in their choice of academic pairing:

$$c - r = cp + (1-p)0$$

giving

$$p = 1 - \frac{r}{c} = 1 - \frac{\rho}{\eta} \equiv p^*$$

at equilibrium. Here players chose to be deviants or pseudos iteratively, with the choices dependent on what they have learnt from previous play.[7] Notice also, the game is actually extended to include the choice of verification or non-verification. Learning assumes that the average scores obtained by deviants and by pseudos in previous rounds are common knowledge. Alternatively, a local iterative learning scheme, where for example each individual keeps a cumulative score of the different choices obtained by one-self in previous rounds, can be envisaged. This can be e.g. in the form of winner-take-all neural networks [von der Malsburg, 1973].

When dealing with that proportion of the deviant that is not checking things out, $(1-q)$, a pseudo can expect a payoff of [*pb*+(1-*p*)*a*]. But, when dealing with those that do check it out, pseudos will sadly have to revert to work with their own kind and just get *a*. So on average, the proportion of deviants who conduct the "check-out your partner" is given by:

$$(1-q)[pb + (1-p)a] + aq = c-r$$

Dividing by *a* and rearranging;

$$(1-q)\left[\xi\left(1 - \frac{\rho}{\eta}\right) + \frac{\rho}{\eta}\right] + q = \eta - \rho$$

and we can then finally arrive at

---

[7] The system dynamics can be of two forms: (i) as new players enter the system, they decide once and for all to be either deviants or pseudos (ii) the system size remains constant while its members decide on the deviant/pseudo choice at every timestep, allowing continued changes of individual states. For the purposes of of this paper, we assume that these two dynamics do not bring significant differences.

$$q = \frac{\alpha - (\eta - \rho)}{\alpha - 1} \equiv q^*$$

where

$$\alpha = \left[\xi\left(1 - \frac{\rho}{\eta}\right) + \frac{\rho}{\eta}\right].$$

So in equilibrium, an average of $q^*$ of deviants will be paying verification costs (or all of them may be doing it $q^*$ of the time). If $q > q^*$, then deviants' population would rise above $p^*$, which as illustrated in Fig. 1, is unstable. And if $q < q^*$, similarly, the population of deviants would start to fall, which is also unstable. In Fig. 1, $q = q^*$ at equilibrium is the point where the deviant's payoff crosses the pseudo's payoff on the vertical line segment between a payoff of $pb$ and $a$, i.e. at the payoff of $c-r$. This point of crossing is $q^*$ of the way down that $c-r$ unit segment, less than the payoff of $[pb + (1-p)a]$ (when $q = 0$) or $1 - q^*$ of the way up from the payoff of $a$ (when $q = 1$). This point of $p = p^*$ and $q = q^*$ stands as a point of stability or an attractor.

The value $\alpha$ is like the weighted (by the ratio of information or verification cost to deviant-deviant payoff, $\rho/\eta$) average of $\xi$ with 1. As $\rho$ changes from 0 to $\eta$, $\alpha$ changes proportionally from $\xi$ to 1. The value for $q$ would then change from $(\xi - \eta)/(\xi - 1)$ to infinity.

The pay-offs to each side is again illustrated in Figure 1. If $p < p^*$, then the right hand side of the deviants' pay-off equation would be less than $c-r$, and so every deviant would learn that it will be much better to pay the check-out cost, identify their academic partner and be guaranteed $c-r$. This would then start to reduce the population of pseudos, so that $p$ would start to rise. If $p > p^*$, on the other hand, deviants would learn to stop checking, and [sometimes] even to turn pseudos, and the population of pseudos would increase so that $p$ begins to decline. In either case, the population would return to its stable state where $p = p^*$. The rates of return are typified by the differences $\Delta$ in the respective payoffs in the following Figure 1.

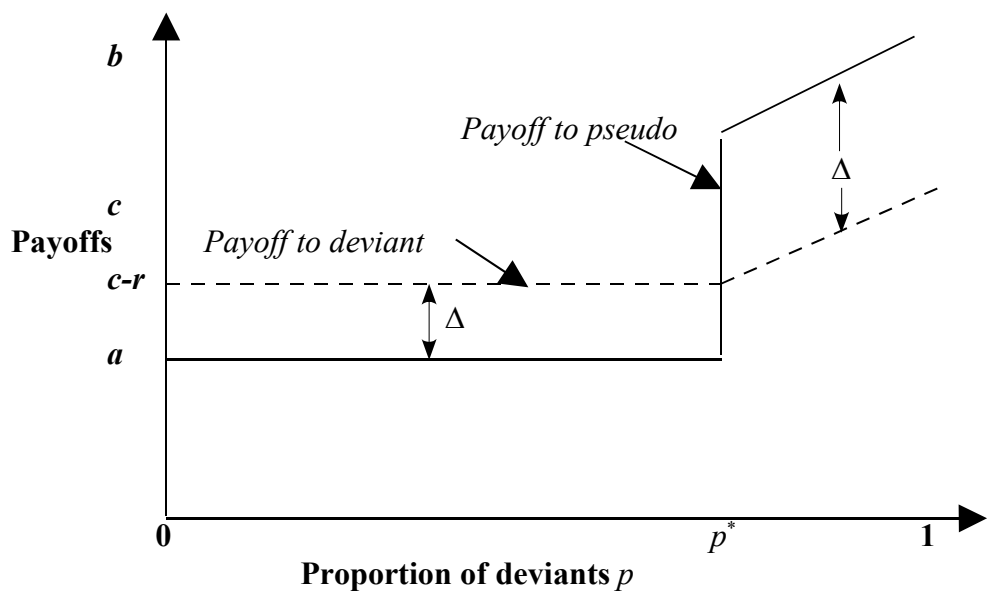

**Figure 1: Payoffs for different deviant proportions in the DD game with information**

From Figure 1, for $p < p^*$, deviants get average payoffs of $c-r$, because they all pay the verification costs, and for $p > p^*$, deviants are paid off as $cp+(1-p)0$, which will increase from $c-r$ to $c$ as $p$ increases from $p^*$ to 1. For pseudos, to the left of $p = p^*$, the average payoff is $a$, since they will be working only with their own kind, but to the right, they are paid off as $bp+(1-p)a$, which increases from $[pb+(1-p)a]$ to $b$ as $p$ increases from $p^*$ to 1. To the right of $p = p^*$, the payoffs to pseudos are greater than that to deviants, so the latter's proportion ($p$) will decrease, and conversely to the left of this point, resulting in the stable point $p = p^*$.[8]

The evolutionary dynamics of the system can thus be represented by the following attractor diagram[9]:

---

[8] What's about this $p = p^*$? If the costs of verification were higher, or if the payoff from deviant working together is lower, then p would be much lower. But we'll follow the deviants' inspiration in this case - How can the payoff for deviants in working together be low? Deviants just perform when they work with their own kind! *Evidennement*!

[9] Notice that the population of pseudos in the institution can vary erratically as soon as too few or too many deviants are opting for the check-out. In equilibrium every deviant is indifferent to choosing either to cooperate or not to cooperate, because they get the same average score for the two cases. However, the choice taken makes a big difference to pseudos!

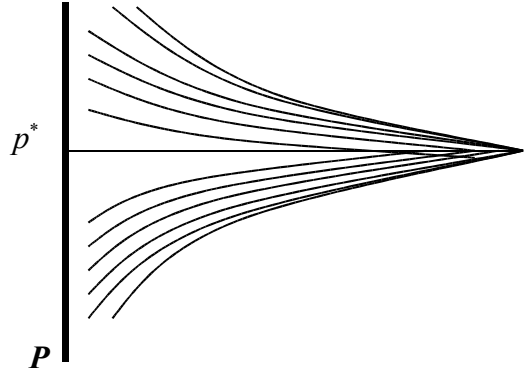

**Figure 2: Attractor diagram for the DD game with information**

## 4. ROLE OF NOISE

Consider introducing noise into the system. In reality, noise is actually required for optimal learning, especially when local learning is implemented. At the early stages, some randomness of choice (despite initial neural network responses) are required to prevent players from becoming immediately stuck to their initial choices and thus not allowing the search for the equilibrium values. In other words, the system needs to "thermalize" in order to reach equilibrium.

Noise in this system would come from errors or uncertainties in payoffs, as well as in information relating to the research partner checking-out process. Let us first look at the effect of noise on the dominance of pseudos in the DD game as well as on the attractor values of $p$ and $q$ in the DD game, firstly with no information and then with information.

With no information, from generalized payoffs we have the pseudo group dominating as long as $\xi - \eta + \frac{(1-p)}{p} > 0$. However, at substantial levels of noise, there would be a non-trivial chance that deviants instead would prosper. If we assume the noise in payoffs to be Gaussian and quantified by its (normalized by $a^2$) variance $\sigma^2$, then the probability for deviants dominating, which equals $p$, is given by the cumulative error function:

$$p = \frac{1}{2}\operatorname{erfc}\left(\frac{\xi - \eta + \frac{(1-p)}{p}}{\sigma/2}\right)$$

where the factor 2 in the argument comes from the quadratic sum of errors in $\xi$ and $\eta$.

For the system with information on prospective partners available through verification, noise in payoffs is expected to smear the equilibrium values of $p$ and $q$ correspondingly, and similar analyses as above could be carried out. What is more relevant here is noise in the information obtained, viz. there is some probability $s$ that the type (deviant or pseudo) of the prospective partner is not really according to the information obtained in the verification process. The average payoff to the deviant after checking is now $(1-s)c$ (where we have not included errors in payoffs). The ESS now requires that

$$(1-s)c - r = cp + (1-p)0$$

giving an equilibrium value of

$$p = 1 - s - \frac{r}{c} = 1 - s - \frac{\rho}{\eta}.$$

Similarly, the equilibrium value of $q$ is now given by

$$(1-q)[pb + (1-p)a]0 + aq = (1-s)c - r$$

which gives

$$q = \frac{\alpha' - [(1-s)\eta - \rho]}{\alpha' - 1}$$

where

$$\alpha' = \xi\left[1 - \left(s + \frac{\rho}{\eta}\right)\right] + \left(s + \frac{\rho}{\eta}\right).$$

The noise has caused the proportion of deviants to drop, and has also changed the proportion of verifiers.

## 5. CONCLUSIONS

We have described and outlined a simple model of an evolutionary game analysing the dilemma of deviants in their conflict with pseudos in the run-up to seniority in some tertiary institutions. The history of deviants and pseudos can also be painted differently bringing a different picture to the set-up to the conflict in the game, payoffs and strategies. Supposedly if in the payoffs we have made pseudo to be equally good at choosing deviant as their partners as deviant themselves are in picking their own, pseudo will then have the evolutionary advantage instead. We have also illustrated the conduct of the associated stability analysis of each group.

To this end we have presented the formal Deviants' Dilemma game. Possibly, the novelty of the game is its one-sidedness, while information coupled with exclusiveness then is seen to overcome the one-sidedness. The outcome of our deviants' dilemma game parallels the findings of Kim and Parker (1995) - the initial advantage points to traditional pseudo academic group-the payoff matrix blatantly favours them, but when information is made available (albeit at cost), exclusiveness of partner-choosing then pushes out this group despite being favoured.

Throughout, the discussion has centred around an academic set-up. Nevertheless, we feel that similar conflicts can be identified in other organisations. We envisage applications to networking of businesses, run-up to seniority in business corporations, and in political organizations. It is also in our interests to explore in future the role that *DD* might play in the emergence of heterogeneity in agent-based models of the economy. . We hope to run extensive computer simulations of the model to verify the analysis done in this paper, and to report the results in a future publication.